\documentclass[runningheads]{llncs}

\usepackage{algorithm,braket,multirow,bigdelim}
\usepackage{algpseudocode,multicol}
\usepackage[table]{xcolor}
\NewDocumentCommand\TODO{ o }{%
    {
        \IfNoValueF {#1} { \texttt{\color{red}[TODO: #1]}}%
\IfNoValueT {#1} { \texttt{\color{red}[TODO]}}%
    }%
}
\usepackage[T1]{fontenc}
\usepackage{graphicx,easyReview,pdfcomment,booktabs,tikz,subcaption,graphicx,mathtools,listingsutf8,tabularx}
\usepackage{listings}
\definecolor{eclipseBlue}{RGB}{42,0.0,255}
\definecolor{eclipseGreen}{RGB}{63,127,95}
\definecolor{eclipsePurple}{RGB}{127,0,85}
\definecolor{lightGray}{HTML}{C1C7CD}
\lstdefinelanguage{sparql}{
	morecomment=[l][]{\#},
	morestring=[b][]\",
	morekeywords={INSERT,DATA,BIND,URI,CONCAT,SELECT,CONSTRUCT,DESCRIBE,ASK,WHERE,FROM,NAMED,PREFIX,BASE,OPTIONAL,FILTER,GRAPH,LIMIT,OFFSET,SERVICE,UNION,EXISTS,NOT,BINDINGS,MINUS,a},
	sensitive=true
}

\lstdefinelanguage{cypher}{
	morekeywords={MATCH,DELETE,RETURN,WHERE,DISTINCT,WITH,CREATE,COUNT,AS,UNION,ALL,is,null,NOT,AND,OR},
	sensitive=true,
	morecomment=[l]{//}, % l is for line comment
}

\usepackage{xparse}
\usepackage{marginnote}
\usepackage{soul}
\usepackage{lipsum}
\usepackage{tikz}
\usetikzlibrary{calc}
\usetikzlibrary{decorations.pathmorphing}

\usepackage[inline]{enumitem}
\newlist{mylist}{enumerate*}{1}
\setlist[mylist]{label=(\roman*)}

\newlength\LineWidth
\newlength\Amplitude
\newlength\SegLength

\definecolor{HLcolor}{RGB}{240,0,0}

\newcommand\tikzmark[1]{%
  \tikz[overlay,remember picture] \node (#1) {};}

\makeatletter
\newcommand{\highlight@DoHighlight}{
  \draw[HLcolor,line width=\LineWidth,decorate,decoration={zigzag,amplitude=\Amplitude,segment length=\SegLength}]  ($(begin highlight)+(0,-2pt)$) -- ($(end highlight)+(0,-2pt)$) ;
}

\newcommand{\highlight@BeginHighlight}{
  \coordinate (begin highlight) at (0,0) ;
}

\newcommand{\highlight@EndHighlight}{
  \coordinate (end highlight) at (0,0) ;
}

\newdimen\highlight@previous
\newdimen\highlight@current

\DeclareRobustCommand*\highlight[1][]{%
  \SOUL@setup
  \def\SOUL@preamble{%
    \begin{tikzpicture}[overlay, remember picture]
      \highlight@BeginHighlight
      \highlight@EndHighlight
    \end{tikzpicture}%
  }%
  \def\SOUL@postamble{%
    \begin{tikzpicture}[overlay, remember picture]
      \highlight@EndHighlight
      \highlight@DoHighlight
    \end{tikzpicture}%
  }%
  \def\SOUL@everyhyphen{%
    \discretionary{%
      \SOUL@setkern\SOUL@hyphkern
      \SOUL@sethyphenchar
      \tikz[overlay, remember picture] \highlight@EndHighlight ;%
    }{%
    }{%
      \SOUL@setkern\SOUL@charkern
    }%
  }%
  \def\SOUL@everyexhyphen##1{%
    \SOUL@setkern\SOUL@hyphkern
    \hbox{##1}%
    \discretionary{%
      \tikz[overlay, remember picture] \highlight@EndHighlight ;%
    }{%
    }{%
      \SOUL@setkern\SOUL@charkern
    }%
  }%
  \def\SOUL@everysyllable{%
    \begin{tikzpicture}[overlay, remember picture]
      \path let \p0 = (begin highlight), \p1 = (0,0) in \pgfextra
        \global\highlight@previous=\y0
        \global\highlight@current =\y1
      \endpgfextra (0,0) ;
      \ifdim\highlight@current < \highlight@previous
        \highlight@DoHighlight
        \highlight@BeginHighlight
      \fi
    \end{tikzpicture}%
    \the\SOUL@syllable
    \tikz[overlay, remember picture] \highlight@EndHighlight ;%
  }%
  \SOUL@
}
\makeatother

\DeclareDocumentCommand\MarkText{O{red}O{1pt}O{5pt}m}{%
  \colorlet{HLcolor}{#1}
  \setlength\Amplitude{#2}%
  \setlength\SegLength{#3}%
  \tikzmark{endquote}\tikzmark{beginquote}\highlight{#4}%
}

\usepackage{glossaries}

\makeglossaries

\newacronym{gql}{GQL}{Graph Query Languages}
\newacronym{gsm}{GSM}{Generalised Semistructured Model}
\newacronym{dag}{DAG}{Direct Acyclic Graph}
\newacronym{llm}{LLM}{Large Language Model}

\definecolor{IBMcyan}{HTML}{82CFFF}
\definecolor{IBMmagenta}{HTML}{FF7EB6}
\definecolor{IBMblue}{HTML}{0F62FE}
\definecolor{IBMpurple}{HTML}{BE95FF}
\newcommand{\vtopo}[1]{\ensuremath{V_\textsf{topo}(#1)}}
\begin{document}
\title{Generalised Graph Grammars for Natural Language Processing}
\author{Oliver R. Fox\orcidID{0009-0005-2483-5672} \and
Giacomo Bergami\orcidID{0000-0002-1844-0851}}
\authorrunning{Fox and Bergami}
\institute{School of Computing, Faculty of Science, Agriculture and Engineering, Newcastle University, Newcastle upon Tyne NE4 5TG, UK
\email{\{O.Fox3,Giacomo.Bergami\}@newcastle.ac.uk}}
\maketitle              % typeset the header of the contribution
\begin{abstract}
%<<<<<<< HEAD
%This seminal paper proposes a new query language for graph matching and rewriting, overcoming some declarativeness limitations of Cypher while outperforming Neo4J on graph matching and rewriting by at least one order of magnitude. This was made possible by exploiting columnar databases (KnoBAB) to represent graphs using the generalised semistructured model.
%=======
This seminal paper proposes a new query language for graph matching and rewriting overcoming {the declarative} limitation of Cypher while outperforming {Neo4j} on graph matching and rewriting by at least one order of magnitude. We exploited columnar databases (KnoBAB) to represent graphs using the Generalised Semistructured Model.
%>>>>>>> 7ea1e0fa7a713b54a3bdcac712abd3a86e931f24

\end{abstract}
%
%
%

%\TODO[8 Pages Max!]

\section{Introduction}
State-of-the-art sentence similarity approaches boil down to %\replace{compute a vector similarity score by computing a vector representation (\textit{embedding}) for each sentence}
{computing a vector representation (\textit{embedding}) for each sentence to determine a similarity score} \cite{reimers-2019-sentence-bert}. This approach does not consider the positionality of some entities within the sentence, and therefore provides wrong results. Furthermore, we also expect such similarity metrics not to be symmetric, we might want to use a similarity to derive how much each sentence implies the second:
%\gls{gql} are becoming increasingly appealing given the recent developments that show the need for the combination of graph data and \gls{llm} \cite{jin2023large}. While the former makes explicit semantic relationships between entities at play, the latter provides rich semantic characterization characterizing the whole ego-net semantics associated with the text. In fact, at present, semantic representations for full sentences as vectors (\textit{embeddings}) for establishing similarities between them for question-answering purposes, despite being currently considered as state-of-the-art solutions \cite{10.1145/3589778}, are particularly harmful because, as the following example illustrates, they do not consider those linguistic relationships.

\begin{example}
Given the sentences: 
\begin{mylist}
\item \label{sa}``\textit{There is no traffic in the Newcastle City {Centre}}'',
                 \item \label{sb}``\textit{Newcastle City {Centre} is trafficked}'',
                 \item \label{sc}``\textit{There is traffic but not in the Newcastle City Centre}'', and
                 \item \label{sd}``\textit{In Newcastle, traffic is flowing}''
\end{mylist}, we expect the similarity between \ref{sc} and \ref{sd} should be very low, as they only agree on the situation within the city centre. In this context, we expect sentence similarity not to be symmetrical, as \ref{sc} or \ref{sd} entail \ref{sa}, but the vice versa should not hold, as part of the information cannot be determined due to missing data; still, the rank of the  {vice versa} should be ranked higher than the one between \ref{sc} and \ref{sd}, as these two sentences contain evidence of conflicting information. Furthermore, \ref{sb} should be significantly dissimilar across all possible sentences, as this is the only sentence referring to traffic appearing within the city centre. 
By representing sentences with vectors using the Sentiment Transformer library \cite{reimers-2019-sentence-bert}, %we observe that sentence similarity is nevertheless symmetric, thus not allowing to capture the aforementioned logical subtlety; furthermore, 
the similarity\footnote{\url{https://osf.io/mvpd2?view_only=f31eda86e7b04ac886734a26cd2ce43d}} across sentences with conflicting information (\ref{sb} and \ref{sc}) is higher than the one between compatible sentences (between \ref{sa} and either \ref{sc} or \ref{sd}), which is undesired. A high similarity between \ref{sa} and \ref{sb} remarks the impossibility of this model to ascertain semantic information depending on the position of specific negation symbols. %The script reproducing such results is freely available. 
\qed
\end{example}

We need to rewrite the sentences first so that two equivalent sentences are rewritten similarly for deriving the embedding. Working under the English language's universal grammar assumption  \cite{Christensen_2019}, we then identify specific grammatical structures for rewriting them using matching and production rules. Given that sentences can be rewritten as a rooted direct acyclic graph while preserving both semantic and syntactic information \cite{DBLP:conf/acl/ManningSBFBM14} (\figurename~\ref{fig:gtbm}) and given that graph query languages postulate the possibility of rewriting a graph into another (\S\ref{sec:rw}), we would then require such an intermediate data processing step for rewriting the sentences under a graph representation{. }{Next, we} can easily derive a \gls{llm} representation \cite{jin2023large}. We would then like to express matching and rewriting patterns independently from the structure of a sentence so that the ways such sentences have to be rewritten would be independent of the sentence structure itself (\figurename~\ref{fig:withresult}). Despite the possibility of doing the following coming from  discrete mathematics literature (\S\ref{ssec:gg}), current graph query languages such as (Open)Cypher %(\S\ref{ssec:oc}) 
are  limited in this regard, as they also require determining how each single pattern's outcome must be combined on the specific sentence structure (\S\ref{sec:wherComplexIs}). This is detrimental, as it doesn't make it possible to fully automate the transformation of the sentences within a syntactically irrelevant representation when semantically similar.

\begin{figure*}[!t]
\centering
\begin{minipage}{0.45\linewidth}
    \begin{subfigure}{\linewidth}
   \subfloat[{Injecting the articles/possessive pronouns ($\lambda$) in $Y$ for an entity $X$ as its own properties, while deleting $\lambda$ and $Y$}]{\label{injprop}
      \includegraphics[width=.7\textwidth]{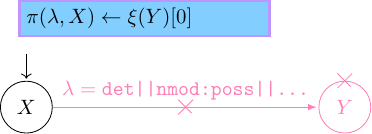}}
    \end{subfigure}
    \begin{subfigure}{\linewidth}
   \subfloat[{Expressing the verb as a binary relationship between subject and direct object}]{\label{binrel}
      \includegraphics[width=.7\textwidth]{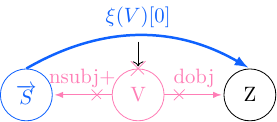}}
    \end{subfigure}
\end{minipage}
    \begin{subfigure}{.45\linewidth}
\subfloat[{Generating an new entity $H'$ coalescing the ones {$\overrightarrow{H}$} under the same conjunction $Z$, while referring to its original constituents via \textbf{orig}.}]{\includegraphics[width=.8\textwidth]{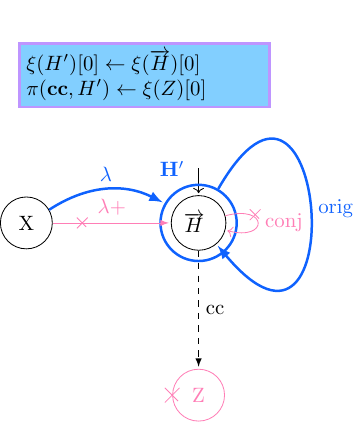}}
    \end{subfigure}
   \caption{Graph grammar production rules \textit{à la GraphLog} in this paper's use case scenario  \cite{graphLog}: thick denotes insertions, crosses deletions, and optional matches are dashed. We extended it with multiple optional edge label matches (${\|}$), key-value association  $\pi(\lambda, X)$ for  property $\lambda$ and node $X$, and multiple node values $\xi(X)$. }\label{bs1}
%\vspace*{-.82cm}
\end{figure*}

\begin{figure*}
\subfloat[{Dependency graph for ``\textit{Alice and Bob play cricket}''}]{\label{fig:gtbm}
\includegraphics[width=.33\textwidth]{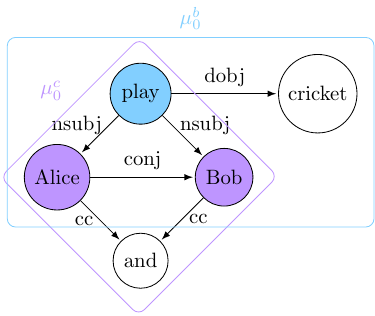}}
\hfill
\subfloat[{Generating a binary relationship between the subject as a single entity and the direct object.}]{\label{fig:resab}
\makebox[7cm][c]{\includegraphics[width=.33\textwidth]{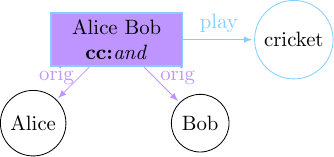}}}\caption{Applying the rewriting rules expressed in \figurename~\ref{bs1}: different colours refer to different graph grammar rules (\textit{b} and \textit{c}), filled nodes in the left (and right) graph refer to the distinct node entry points (and newly generated components).}\label{fig:withresult}
%\vspace*{-.8cm}
\end{figure*}

To overcome these limitations, we propose a new graph data query language that {alleviates} the shortcomings above in Cypher 
by assuming the graphs' acyclicity; it is now possible to visit the graph in reverse topological order, thus starting from the most nested part of the sentence (e.g., subordinate) towards the most apical elements, i.e. its root (verb of the main clause, if not a noun) \cite{info15010034}, while applying the sentence rewriting accordingly. 
We also show that implementing this query language on top of a tailored relational engine for {a} \gls{dag}s\footnote{\url{https://github.com/datagram-db/datagram-db/}} \cite{generalisedGiacomo} outperforms the execution of similar queries over {Neo4j} natively supporting Cypher. All the graphs and queries for our preliminary benchmarks are freely available online\footnote{\url{https://osf.io/btjqw/?view\_only=f31eda86e7b04ac886734a26cd2ce43d}}{.}

\section{Related Works}\label{sec:rw}\label{ssec:gg}
\begin{figure*}[!t]
\begin{minipage}[t]{.30\linewidth}
\subfloat[{Graph $g$ to be mathed}]{\label{gdb1}
\includegraphics[width=.8\textwidth]{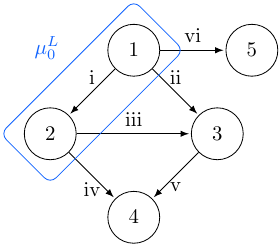}}
\end{minipage}
\begin{minipage}[t]{.14\linewidth}
\subfloat[{Graph\\ pattern $L$}]{\makebox[2cm][c]{\label{gq1}
\includegraphics[width=.45\textwidth]{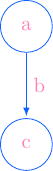}}\\\quad}
\end{minipage}
\begin{minipage}[t]{.49\linewidth}
\subfloat[{Morphism table $M[L,g]$ where each row describes a morphism $\mu_i$ between the graph matching $L$ and the graph $g$.}]{\label{mt1}
		\makebox[5.5cm][c]{\resizebox{.23\linewidth}{!}{\begin{tabular}{|>{\columncolor{lightGray}}c|c|c|c|}
			\hline
			\rowcolor{lightGray}
			\textbf{} & \textbf{\color{IBMmagenta}a} & \textbf{\color{IBMmagenta}b} & \textbf{\color{IBMmagenta}c}\\
			\hline
			$\boldsymbol{\color{IBMblue}\mu_0^L}$ & {\color{IBMblue}1} & {\color{IBMblue}i} & {\color{IBMblue}2} \\
			\hline
			$\boldsymbol{\mu_1^L}$ & 1 & ii & 3 \\
			\hline
			$\boldsymbol{\mu_2^L}$ & 2 & iii & 3 \\
			\hline
			$\boldsymbol{\mu_3^L}$ & 2 & iv & 4 \\
			\hline
			$\boldsymbol{\mu_4^L}$ & 3 & v & 4 \\
			\hline
			$\boldsymbol{\mu_5^L}$ & 1 & v & 5 \\
			\hline
		\end{tabular}}}}
\end{minipage}

   \caption{Listing all the subgraphs of $g$ being a solution of the subgraph isomorphism problem of $g$ over $L$.}\label{subgraphNoData}
%\vspace*{-.8cm}
\end{figure*}

{Graph grammars \cite{rozenberg} are to be considered the theoretical foundations of  {current} graph query languages, as they express the capability of \textit{matching} specific patterns $L$ \cite{perez2009} within the data through reachability queries while applying modifications to the underlying graph database structure (\textit{graph rewriting}) $R$, thus producing a single graph grammar production rule $L\xrightarrow{f} R$, where there is an implicit morphism between some of the nodes (and edges) matched in $L$ and the ones appearing in $R$: the nodes (and edges) only appearing in $R$ are considered as newly inserted nodes, while the nodes (and edges) only appearing in $L$ are considered as removed edges; we preserve the remaining matched nodes.} {Each rule is then considered}  as a function {$f$}, taking a graph database as {an} input and returning a transformed graph. 
%\begin{figure*}[!t]
%\centering
%%\begin{minipage}[t]{.45\linewidth}
%
%
%\subfloat[{Formulating $L$ from \figurename~\ref{gdb1}.}]{\label{cypherQuery}
%		\makebox[5.5cm][c]{\lstinline[language=cypher]|MATCH (a)-[b]->(c)|}}\subfloat[{Morphism table resulting from \subref{cypherQuery}.}]{\label{table:simpleMatchCypher}
%		{
%		\resizebox{.45\textwidth}{!}{
%		\begin{tabular}{|l|l|l|}
%			\hline
%			\rowcolor{lightGray}
%			\textbf{a} & \textbf{b} & \textbf{c} \\
%			\hline
%			\texttt{(\{name: "play"\})} & \texttt{[:dobj]} & \texttt{(\{name: "cricket"\})} \\
%			\hline
%			\texttt{(\{name: "play"\})} & \texttt{[:subj]} & \texttt{(\{name: "Alice"\})} \\
%			\hline
%			\texttt{(\{name: "play"\})} & \texttt{[:subj]} & \texttt{(\{name: "Bob"\})} \\
%			\hline
%			\texttt{(\{name: "Alice"\})} & \texttt{[:conj]} & \texttt{(\{name: "Bob"\})} \\
%			\hline
%			\texttt{(\{name: "Alice"\})} & \texttt{[:cc]} & \texttt{(\{name: "and"\})} \\
%			\hline
%			\texttt{(\{name: "Bob"\})} & \texttt{[:cc]} & \texttt{(\{name: "and"\})} \\
%			\hline
%		\end{tabular}}}}
%%\end{minipage}
%\caption{Cypher representation of \figurename~\ref{subgraphNoData}.}
%\vspace*{-.5cm}
%\end{figure*}
The process of matching {$L$} is usually expressed in terms of subgraph isomorphism{:} {given} two graphs $G$ and $L$, {we} determine whether $G$ contains a subgraph $G_i$ that is isomorphic to $L${, i.e. there is a bijective correspondence $L\xleftrightarrow{\mu_i} G_0$ between the nodes and edges of $L$ and $G_i$}.  {In graph query languages, we are then considering $G$ as our graph database and returning  $f(G_i)$ for each matched subgraph $G_i$}. {When no rewriting is considered, each possible match $G_0$ for $L$ is usually represented in a tabular form \cite{perez2009,10.1145/3183713.3183724}, where the column header provide{s} the node and edge identifiers (e.g., variables) $j$ from $L$, each row reflects each matched graph $G_i$, and each cell corresponding to the column $j$ represents $\mu_i(j)$. \tablename~\ref{mt1} provides a graphical depiction for this usual representation of graph morphisms.} 
\figurename~\ref{bs1} illustrates some graphical representation of graph grammar rules as defined for GraphLog for transforming the dependency graph into a more compact graph representation: we can first create the new nodes required in $R$, while updating or removing $x$ as determined by the node or edge $f(\mu_i^{-1}(x))$ occurring in $R$. Deletions can be performed as the last operations from $R$. GraphLog still allows running of one single grammar rule at a time, while authors assume to have a generic graph. Having a DAG is a strict requirement in our scenario considering a graph grammar with multiple rules: about \figurename~\ref{fig:withresult}, it is deemed appropriate to apply the rules starting from the lower nodes towards the upper ones.

\section{Cypher's Limitations and Proposed Query Language}\label{sec:wherComplexIs} Cypher suffers from the limitations posed by the property graph data model which, by having no direct way to refer to the matched nodes or edges by reference, forces the querying user to always refer to the properties associated to them; as a consequence, the resulting morphism tables are carrying out redundant information that cannot reap the efficient data model posed by columnar databases, where entire records can be reference{d} by their {ID}. This is evident for \lstinline[language=cypher]|DELETE| statements, voiding objects represented within the morphisms. This limitation of the property graph model, jointly with the need for representing acyclic graphs, motivates us to use the \gls{gsm} as an underlying data model for representing graphs, thus allowing us to refer to the nodes and edges by their {ID} \cite{generalisedGiacomo}. Consequently, our implementation represents morphisms for acyclic property graphs as per \figurename~\ref{mt1}.

The current {Neo4j} implementation does not support the theorised graph incremental views for Cypher \cite{10.1145/3183713.3183724}. At the same time, it is not possible to entirely create a new graph without restructuring or expanding a previously loaded one; returning a new graph and rewriting a previous match will come at the cost of either restructuring the previously loaded graph, thus requiring additional overhead costs for re-indexing and updating the database while querying, or by creating a new distinct connected component within the loaded graph. As it is impossible to refer by the nodes and edges through their {ID,} thus exploiting graph provenance techniques for mapping the newly created nodes to the ones from the previously loaded graph \cite{10.14778/3436905.3436911},  we are therefore forced to join the loaded nodes with the newly created ones repeatedly. Our proposed approach avoids such cost via the aforementioned morphism representation while keeping track of the restructuring operations (property update, node insertion, deletion, and substitution) over a graph $g$ within an incremental view $\Delta(g)$.

Cypher does not ensure to apply the graph rewriting rules as intended in our scenarios: let us consider the dependency graph generated from the recursive sentence ``\textit{Matt and Tray believe that either Alice and Bob and Carl play cricket or Carl and Dan will not have a way to amuse themselves}'' and let us try to express patterns \subref{binrel} and \subref{rev_sol} as two distinct \lstinline[language=cypher]|MATCH|-es with their respective update operations: we observe that, instead of generating one single connected component representing the result, we will generate as many distinct connected components as subgraphs being identified as matching the patterns, while this does not occur with a simple sentence structure where we achieve the correct result as in \figurename~\ref{fig:withresult}. {We must \lstinline[language=cypher]|MATCH| elements of the graph multiple times, constantly rejoining on data previously \lstinline[language=cypher]|MATCH|-ed in earlier stages of the query.} %\pdfcomment[author=Ollie]{In reference to what we were referring to when talking about constantly rejoining, like with grouping Alice Bob Carl, Carl Dan, and matching again for Alice Bob Carl or Carl Dan. You've outlined this already but added in this sentence for extra clarity if you think it fits.} 
This then postulates the inability of such language to automatically apply an order of visit for restructuring the loaded graph while not expressing an automated way to merge each distinct transformed graph into one cohesive, connected component. This then forces the expression of a generic graph matching and rewriting mechanism to be dependent on the specific recursive structure of the data{. Thus,} requiring the creation of a broader query, where we need to explicitly instruct the query language on the correct way to visit the data while instructing how to reconcile each generated subgraph from each morphism within one final graph.
 
During the delineation of the final Cypher query succeeding in obtaining the correct rewritten graph, we also highlighted the impossibility of Cypher to propagate the temporary result generated by a rewriting rule and propagate it to another rule to be applied upstream: this requires carrying out intermediate sub-queries establishing connections across patterns sharing intermediate nodes, as well as the re-computation of the same intermediate solutions,  such as node grouping. Since Cypher also does not support the explicit grouping of nodes based on a pattern as in \cite{DBLP:conf/btw/JunghannsPR17}, this required us to identify the nodes satisfying each specific pattern, label them appropriately in a unique way, and then compare the result obtained.  We show this limitation can be overcome by providing two innovations: first, using nested relational tables for representing morphisms, where each nest will contain the sub-pattern of interest possibly to be grouped. Second, we track any node substitution for entry-point node matches via incremental views. This substitution can be easily propagated at any level by considering the transitive closure of the substitution function, while the order of visit {in} the graph guarantees the correctness of the application of such substitution.

{The Cypher query constructed for the specific matches referring to the sentence "\textit{Matt and Tray...}", will not fully execute on a different sentence without the given dependencies, as no match is found, and therefore no rewriting can occur. Current graph query languages are meant to return a subgraph from the given patterns. In Cypher, you must abide with what is contained within the data, if the data is not there we need to remove the match from the query, which we cannot forecast in advance. This results in constant analysis of the data. For us the intention is to have graph grammar rewriting rules whereby if a match is not made, no rewriting occurs.}

By leveraging such limitations of {Cypher} while juxtaposing the desired behaviour of the language, we derive a declarative graph query language where patterns can be expressed similarly to \figurename~\ref{fig:withresult}. 
Due to the lack of space, we refer to our wiki\footnote{\url{https://github.com/datagram-db/gsm_gsql/wiki/Syntax}} for a complete reference for the syntax of our language. 

\section{Proposed Solution}\label{sec:ql}
%\paragraph*{Query Language} \TODO 

\paragraph*{Physical Storage.} We represent a graph database as a collection of graphs $G$, where each graph is defined according to the \gls{gsm} \cite{generalisedGiacomo}, where each node is a semistructured object. Each edge is a labelled containment relationship between objects:
each node $v$ of a graph $g\in G$ has labels ($\ell(v)$) and values ($\xi(v)$) vector, both loaded in dedicated tables for fast retrieval. The physical model reflects the one of KnoBAB \cite{info15010034}: to match a node by $\ell$, each node is loaded as a tuple $\braket{\ell(u),g,u}$ in an \textsf{ActivityTable} at a specific offset \texttt{off}; non-null key-value associations for keys $k$ are stored as a record $\braket{g,v,\texttt{off}}$ in a \textsf{AttributeTable}\textsubscript{k}; each edge $u\xrightarrow{\ell}v$ with {ID} $e$  and label $\lambda$ in $g$ is represented as a record $\braket{\ell(u),g,u,e,v}$  stored in a table \textsf{PhiTable}\textsubscript{$\lambda$}.
In addition, we define an incremental view over the graph database $\Delta(g)$, which will store the update information referring to the running of the operations listed in $R$. This is detailed in the next paragraph.

\paragraph*{Implementing the Query Algorithm.}
First, load each acyclic dependency graph $g$ in primary memory and index them within the physical model described in the previous subsection. At indexing time, we create the primary and secondary index for each table \cite{info15010034} while topologically sorting their vertices in $\vtopo{g}$ \cite{generalisedGiacomo}. 

Second, we parse the query of choice and we rewrite it into an internal representation; we ensure the minimisation of the data access by running each query pattern occurring across the graph grammar only once{,} while reusing the same result multiple times; the results are stored in a relational table, {in which the headers refer} to the node and edge variables provided within each matching graph. We separate the optional patterns from the required ones. After this, we merge the intermediate edges {, similar} to SPARQL semantics \cite{perez2009}: {we} represent each graph matching $L$ as an equi-join query between all the previously-instantiated tables and, if there are any optional matches to consider, we compute left-joins between the outcome of such an equi-join. Then, we start nesting the morphisms which, on the other hand, are supported in neither Cypher nor SPARQL: for each aggregated node $\overrightarrow{H}$ associated to an incoming edge, we perform a {\textit{group by}} over the variable nodes not appearing as descendants of $\overrightarrow{H}$ while we nest cells referring to descendant nodes and edges within a nested relationship. After this, we instantiate all the nested morphism tables for each matching pattern of interest. We associate each morphism table $M[g,L]$ a primary blocked index referring to the entry point of the match as declared within each graph match query.
%\begin{algorithm}[!t]
%
%   % \begin{multicols}{2}
%\caption{Rewriting phase}
%\begin{algorithmic}
%\ForAll{$g\in G$}
%\State $\Delta(g)\gets$\textbf{new} GraphView($|V(g)|$)
%\ForAll{$v\in \vtopo{g}$ \textbf{s.t.} $v\notin \Delta(g).$deleted}
%\ForAll{$L_\Theta\to R\in \mathcal{G}$ \textbf{s.t.} $M[L,g]\neq\emptyset$}
%\ForAll{$\mu_v^i\in M[L,g]$ \textbf{s.t.} $\forall \texttt{col}. \neg\textsf{Optional}_L(\texttt{col})\Rightarrow \mu_v^i(\texttt{col})\notin \Delta(g).$deleted}
%\State \algorithmicif\; $\Theta\neq\textbf{true}$ \textbf{and not} $\Theta(\mu_v^i)$ \algorithmicthen\; \textbf{continue} 
%\ForAll{\texttt{op}$\in R.$operations}
%\If{\texttt{op}$=u^+$}  $\mu_v^i(u)\gets\Delta(g).$newObject
%\ElsIf{\texttt{op}$=u^-$ \textbf{or} \texttt{op}$=e^-$ } $\Delta(g).$deleted.add($\mu_v^i(u,e)$)
%\Else\, \textbf{update with} $R.$operation$(\mu_v^i)$
%\EndIf 
%\EndFor
%\EndFor
%\EndFor
%\EndFor
%\EndFor
%\end{algorithmic}
%%\end{multicols}
%
%\end{algorithm}

Third, we apply the rewriting query for each graph $g$: we visit the reverse $\vtopo{g}$ while retaining the nodes appearing in the primary index of a non-empty morphism table $M[g,L]$ for each production rule $L_\Theta\to R\in\mathcal{G}$: we skip the associated morphisms if either a previously matched node was deleted and not replaced with a new one, or if it does not satisfy a possible \texttt{WHERE} condition $\Theta$ associated to $L_\Theta$. For the remaining morphisms, we run the operations listed in $R$ in order of appearance: for each \texttt{new \textbf{x}} operations, we temporarily associate a newly generated node from $\Delta(g)$, which will store a mini-database $\Delta(g)$.db for newly created objects, to a variable \texttt{x}; for objects updating their label, properties, or values, we also keep track of such changes within $\Delta(g)$.db; {further} explicitly keep{ing} track of deleted nodes in $\Delta(g)$.deleted, thus discarding the evaluation of any subsequent pattern that originally contained such a previously matched node. Query entry-point nodes $u$ being deleted and then replaced by newly inserted nodes $v$ are tracked through a replacement relationship $\Delta(g).\Re$, while removing $v$ from the previous set of removed objects.

Last, we return {the final graph to the user} by merging the incremental graph updates stored in  $\Delta(g)$ with the original graph loaded in primary memory $g$ and returning this to the querying user, thus providing an example of late materialisation.
Due to page limitations, we resort to the description of the whole algorithm to future works.

	\begin{table}[!t]
		\centering
		\resizebox{.8\textwidth}{!}{\begin{tabular}{ll|p{3cm}|p{1.5cm}|p{2.5cm}|p{1.5cm}}
			\toprule
			%\rowcolor{lightGray}
				
				\multicolumn{2}{c|}{\textbf{Data Model}}  & \textbf{Loading/Indexing (avg. ms)} & \textbf{Querying (avg. ms)} & \textbf{\vtop{\hbox{\strut Materialisation}\hbox{\strut (avg. ms)}}} & \textbf{Total (ms)}\\
\midrule
				
				\multirow{2}{*}{Neo4J} \rdelim\{{2}{2mm} & Simple & $2.33\cdot 10^0$ & $1.33\cdot 10^1$ & N/A & $1.57\cdot 10^1$ \\
				
				& Complex & $4.00 \cdot 10^0$& $5.20\cdot 10^1$ & N/A & $5.60\cdot 10^1$ \\
				
				\multirow{2}{*}{GSM} \rdelim\{{2}{0mm} & Simple & $2.32\cdot 10^{-1}$ & $1.22\cdot 10^0$ & $4.78\cdot 10^{-2}$ & $1.50\cdot 10^0$ \\
				
				& Complex & $6.91\cdot10^{-1}$ & $2.10\cdot 10^0$ & $1.60\cdot 10^{-1}$ & $2.95\cdot 10^0$ \\
				
			\bottomrule
		\end{tabular}}
		\caption{\label{table:executionTimeOfQueries}Table displaying results from rewriting the aforementioned sentences. }
		%\vspace{-30pt}
	\end{table}
\subsection{Empirical Evaluation}\label{sec:ee}

%Our preliminary results were run over a \TODO[Oliver's PC Specifications]. 
%\pdfcomment[author=Ollie]{If specifications are still needed, they are: Intel® Core™ i7-8750H CPU @ 2.20GHz × 12, 32GB RAM, GPU NVIDIA 1050ti, Ubuntu 22.04.3. Not sure if you want to add this in within the paragraph or as a footnote?}
We considered two distinct dependency graphs, the one in \figurename~\ref{fig:gtbm} and the one resulting from the dependency parsing of the ``\textit{Matt and {Tray}\dots}'' sentence from \ref{sec:wherComplexIs}. We loaded them in both Neo4{j} and our proposed GSM database. In Cypher, we then run the query as formulated in the previous section, while we {construct} a fully declarative query in our proposed graph query language syntax directly representing an extension of the patterns in \figurename~\ref{bs1}. 
Examining Table \ref{table:executionTimeOfQueries}, we can see %with %the complex query\add{,} the range in execution times and inefficient performance, as 
our solution consistently outperforms {the} Neo4j solution by one order of magnitude. Furthermore, the data materialisation phase does not significantly impact the overall running time, as its running times are always negligible compared to the other ones. {Additionally,} Neo4{j} does not consider a materialisation phase, as the graph resulting from the graph rewriting pattern is immediately returned and stored as a distinct connected component of the previously loaded graph. %\pdfcomment[author=Giacomo]{You are missing the whole set of motivations: this should be better sketched by comparing the two approaches together.}
This then clearly remarks the benefit of the proposed approach for rewriting complex sentences into a more compact machine representation of the dependency graphs. %\textit{This motivates us to investigate further the presence of further grammatical patterns within the data to obtain homogeneous graph representations of sentences expressing similar or compatible concepts, thus overcoming the aforementioned limitations within vector-based sentence similarity. }

%\remove{Concerning Neo4j and OpenCypher, running the larger query was temperamental, only returning the expected outcome sometimes. If the query is split up into smaller segments, then the query executes as expected, however once bundled together as one query the graph would return something undesired, and not always run the last section of the query. } \pdfcomment[author=Ollie]{After re-creating the large query this behaviour ceased.} % This overall undermines the trustworthiness of the aforementioned graph database for carrying out multi-step graph query rewriting tasks.

\section{Conclusion and Future Works}
This paper starts to address the problem with current solutions for sentence similarities; we postulate that this can be solved by first rewriting the sentences according to their semantics by underlying grammar rules of English when expressed as dependency graphs. For this paper, we rewrite sentences expressed in dependency graphs, for which we designed a novel query language that is more efficient than state-of-the-art graph databases and provides better declarative support. Motivated by the current experiments, future works will investigate the option of additional grammatical rules for rewriting the sentences, as well as provide better scalability analyses.
Acyclic graphs are commonly used in many other contexts, such as citation networks for bibliography \cite{10.1093/comnet/cnu039} or taxonomical representation of entities, from which we can conveniently derive vectorial representation for such entities. Furthermore, we can always freely represent generic graphs \gls{dag}. Future works will also contextualise this on these domains.
\bibliographystyle{splncs04}
\bibliography{refs}

\end{document}